\documentclass[5p]{elsarticle}
\usepackage{graphicx}% Include figure files
\usepackage{graphics}% Include figure files
\usepackage{dcolumn}% Align table columns on decimal point
\usepackage{bm}% bold math
\usepackage{chemformula} % \ch{} chemformula 
\usepackage{siunitx}
\usepackage{caption}
\usepackage{subcaption}
\newcommand{\syb}{SrYb$_{2}$O$_4$}
 
\newcommand{\sdy}{SrDy$_{2}$O$_4$}

\newcommand{\sln}{Sr$Ln_{2}$O$_4$}
\newcommand{\sho}{SrHo$_{2}$O$_4$}
\newcommand{\stb}{SrTb$_{2}$O$_4$}
\newcommand{\stm}{SrTm$_{2}$O$_4$}

\newcommand{\tm}{Tm$^{3+}$}
\newcommand{\lnt}{$Ln^{3+}$}

\newcommand{\mpsr}{$\mu^+$SR}
\begin{document}
%\preprint{}

\title{Crystal field effects in the zig-zag chain compound \stm}

\author{A. Bhat Kademane$^{1}$}
\ead{abhijit.bhatkademane@uis.no}
\author{D. L. Quintero-Castro$^{1,2}$}
\ead{diana.l.quintero@uis.no}
\author{K. Siemensmeyer$^{2}$}
\author{\\\ C. Salazar-Mejia$^{3}$}
\author{D. Gorbunov$^{3}$}
\author{J. R. Stewart$^{4}$}
\author{H. Luetkens$^5$}
\author{C. Baines$^5$}
\author{Haifeng Li$^{6,7}$}

\address{$^{1}$University of Stavanger, 4021 Stavanger, Norway.\\
$^{2}$Helmholtz Zentrum Berlin f\"ur Materialien und Energie, D-14109 Berlin, Germany. \\
%$^{3}$ Technical University of Denmark, 2800 Kgs. Lyngby \\
$^{3}$Hochfeld-Magnetlabor Dresden (HLD-EMFL), Helmholtz-Zentrum Dresden-Rossendorf, 01328 Dresden, Germany.\\
$^{4}$ISIS Neutron and Muon Source, Rutherford Appleton Laboratory, Didcot, OX11 0QX, UK. \\
%$^{5}$Jülich Center for Neutron Science, 52428 Jülich, Germany. \\
$^5$Paul Scherrer Institut, 5232 Villigen PSI, Switzerland.\\
$^6$Jülich Center for Neutron Science,52428 Jülich, Germany.\\
$^7$Joint Key Laboratory of the Ministry of Education, Institute of Applied Physics and Materials Engineering, University of Macau, Avenida da Universidade, Taipa, Macao SAR 999078, China.
}

\date{\today}

\begin{abstract}

The single ion properties of the zig-zag chain compound \stm\ have been investigated using heat capacity, magnetic susceptibility, magnetization, inelastic neutron scattering, and polarized muon spectroscopy. Two crystal field models are employed to estimate the single ion properties; a Density Function Theory based model and an effective charge model based on the Hutchings point charge model. The latter describes our experimental results well. This model estimates an easy-axis anisotropy for one of the \tm\ sites and an easy-plane anisotropy for the second site. It also predicts a mixed ground state with dominating $J = 0$ characteristics for both sites. Additionally, muon spin rotation/relaxation (\mpsr) spectra reveal oscillations, typically a sign of long-range magnetic order.  However, the temperature dependence of the precession frequency and the relaxation rates indicate that the system is in an extended critical regime and the observed relaxation is actually dynamic.

\end{abstract}

%\pacs{75.25.-j, 75.40.−s, 75.50.−y, 75.30.Gw, 75.40.-s}    

\maketitle

\section{Introduction}
It was previously thought that rare-earth-based magnetic systems were well described in terms of classical long-range order of the total angular momentum and that unconventional magnetic phases were only realized in pure spin systems. However, that picture is rapidly changing, as it has been shown that crystal field effects together with magnetic frustration, low coordination, dipolar interactions, and low dimensionality can cast the perfect ground for exotic phenomena in rare-earth systems, such as cooperative paramagnetism \cite{Gardner1999}, potential spin liquid phases \cite{Shen2016,Chillal2020}, noncollinear order \cite{Irkhin2003} and dimerization \cite{Hara2012, Hester2019}.

\begin{figure*}[htb!]
\centering
        \includegraphics[width=1\textwidth]{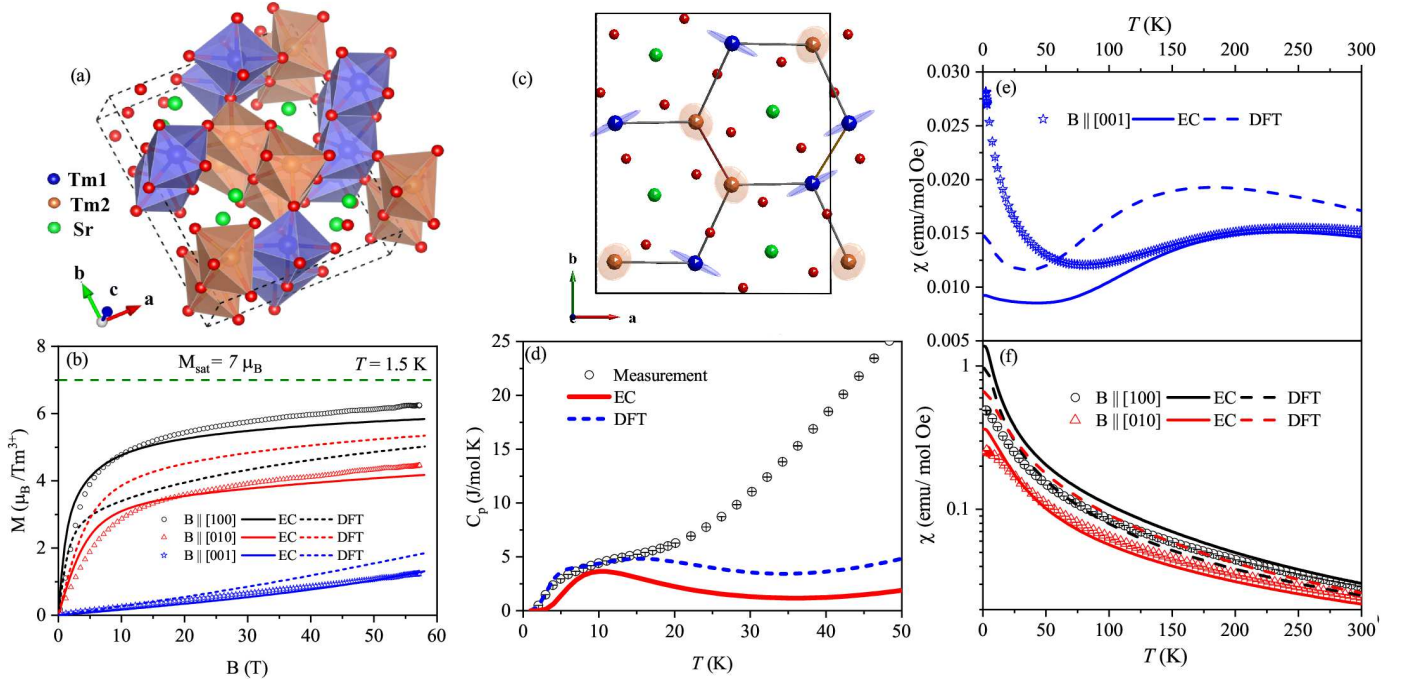}
         \caption{(Color online) (a) \stm \ unit cell. Sr$^{2+}$ ions are shown in green, O$^{2-}$ in red, Tm1 in blue, and Tm2 in orange. \tm \ zig-zag chains run along the $c$-axis and form a distorted honeycomb in $ab$-projection. (b) Measured magnetization (symbols) as a function of the magnetic field at 1.5\,K, with magnetic fields applied along the main crystallographic axes. Calculated magnetization using the EC model (straight lines) and the DFT model (dashed lines). (c) Visual representations of g-tensor ellipsoid for Tm1 and Tm2 for the EC model. Tm1 shows easy-axis anisotropy while Tm2 shows easy-plane anisotropy. There is no component along the $c$-axis for both ions. (d) Low-temperature heat capacity of \stm (data open circles). The blue dashed line corresponds to the calculated Schottky anomaly using the DFT model, and the red solid line corresponds to the EC model. (e,f) Temperature dependence of DC magnetic susceptibility of \stm \ (data shown as symbols) with a magnetic field of 0.1\,T applied along (e) the $c$-axis and (f) $a$- and $b$-axes. The figures show the calculated susceptibility using the DFT model as dashed lines and the EC model as solid lines. } 
    \label{FIG:Bulk}
\end{figure*}

Unconventional magnetic phenomena have been reported for several members of the \sln \, family of compounds, where \emph{Ln} are rare earth ions \cite{Karunadasa2005,Petrenko2014,Qureshi2021,Qureshi2021_2,Gauthier2017,Young2019}. These extend from long-range incommensurate structures in \stb \, \cite{Li2014}, coexisting distinctive types of short-range orders in \sho\, \cite{Fennell2014,Wen2015} and \sdy \cite{Fennell2014,Gauthier2017_2}, and coexisting noncollinear long-range and short-range order in \syb \, \cite{Quintero2012}. These compounds crystallize in the orthorhombic space group 62.\emph{Pnam}, where two crystallographically inequivalent trivalent rare earth ions are surrounded by distorted oxygen octahedra, with monoclinic $C_{s}$ site symmetry, and form two zig-zag chains running along the $c$-axis (Fig. \ref{FIG:Bulk}(a)).

The low-lying crystal field scheme plays a key role in forming highly anisotropic magnetic properties in  \sln \ family \cite{Petrenko2014}. However, determining the crystal field scheme has been a non-trivial task due to the two inequivalent \lnt in low symmetry environments. Nevertheless, successful modeling of the crystal field schemes for some members of this family has been reported \cite{Fennell2014,Malkin2015}.

In \stm \, \tm \ has electronic configuration [Xe] 4$f^{12}$ (non-Kramers ion). A $13$- fold degenerate ground state is expected for the lowest energy multiplet $^{3}H_{6}$ ($S$= 1, $L$ = 5, and $J$ = 6) for every \tm. crystal field only ground state is usually a singlet for the systems involving non-Kramers rare earth ions, and there is no magnetic ordering down to the lowest temperatures \cite{Cooper1971}. This is in agreement with the results reported by Haifeng Li $et.$ $al$. \cite{Li2015}, which shows the absence of short and long-range magnetic ordering in \stm \, down to 60\,mK. 

 Here, we present two aspects related to the single ion effects in \stm. In the first part, we report the results of our magnetic characterization through heat capacity, magnetization, magnetic susceptibility, inelastic neutron scattering (INS) measurements, and we implement two crystal field models. In the second part, we present the observation of quasistatic order observed with muon spin rotation/relaxation (\mpsr).

The standard method used in determining crystal fields involves searching for crystal field parameters (CFP) in parameter space that fit thermodynamic and spectroscopic properties. This procedure has multiple challenges. The major challenge is, degenerate sets of CFPs are expected for a low-symmetry non-Kramers system with multiple magnetic ions. To overcome these challenges, the crystal field problem in \stm\ was approached with an \emph{ab-initio}  Density Functional Theory based model (DFT model) \cite{Novak2013} and effective charge model (EC model) based on the point charge model (PC model) \cite{Dun2021}.

The DFT model is based on methodology proposed by Nov{\'a}k $et.al$ \cite{Novak2013}. In this method CFPs are obtained by using maximally localized Wannier functions (MWLF) with an all-electron DFT implementation. The advantage of this method is that, there is only one parameter that needs to be determined called `charge transfer energy', $\Delta$. The `charge transfer energy' is estimated by,
\begin{equation}
  \Delta \cong E_{tot}(4f^{(n+1)}, N_{val}-1) - E_{tot}(4f^{(n)},N_{val})  
\end{equation}
where $n$-is number of electrons in 4f shell, $N_{val}$ is number of electrons in the valence band, $E_{tot}(4f^{(n)},N_{val})$ is the ground state total energy, and $E_{tot}(4f^{(n+1)},N_{val}-1)$ is the excited state energy. This method has been successfully implemented on several rare earth systems \cite{Novak2013,Novak2015,Novak2014}. 

The EC model is based on Hutchings crystallographic PC model \cite{Hutchings1964}. In this method, the crystallographic PC model was modified to account for oxygen's point charge with adjustable effective charge and effective radius. In our case, the crystallographic PC model with naive use of standard charge (i.e. $-2e$ for oxygen) and crystallographic coordinates at the position of the nucleus of oxygens did not result in an accurate prediction of the magnetic properties. Thus, oxygen charges and their displacement from crystallographic positions are fitted to INS spectra. This approach is more realistic compared to the crystallographic PC model and unlike the standard CFP fit approach avoids overparameterization of fitted parameters. Similar semi-empirical improvisation has been successfully implemented for single-molecule magnets (SMM) \cite{Baldovi2012} and for rare-earth pyrochlores \cite{Dun2021}.
%where one electron was transferred in

In the second part, we present \mpsr\ results. Typical static order in magnetic systems with zero fields \mpsr \ (ZF-\mpsr) manifests as oscillations in the asymmetry spectra. These oscillations are due to muon precessing in the internal magnetic field associated with magnetically ordered surroundings \cite{Yaouanc2011}. Similar observations in \stm\, contradict the absence of magnetic order reported in Ref. \cite{Li2015}. Here, we prove that the quasistatic order observed in \mpsr \ is due to a muon induced lattice distortion. In this work, we use \emph{ab-initio} techniques to identify the muon stopping site and qualitatively discuss the impact of muon implantation on the crystal field with the aid of the PC model.

\begin{figure*}[htb!]
\centering
       \includegraphics[width=1\textwidth]{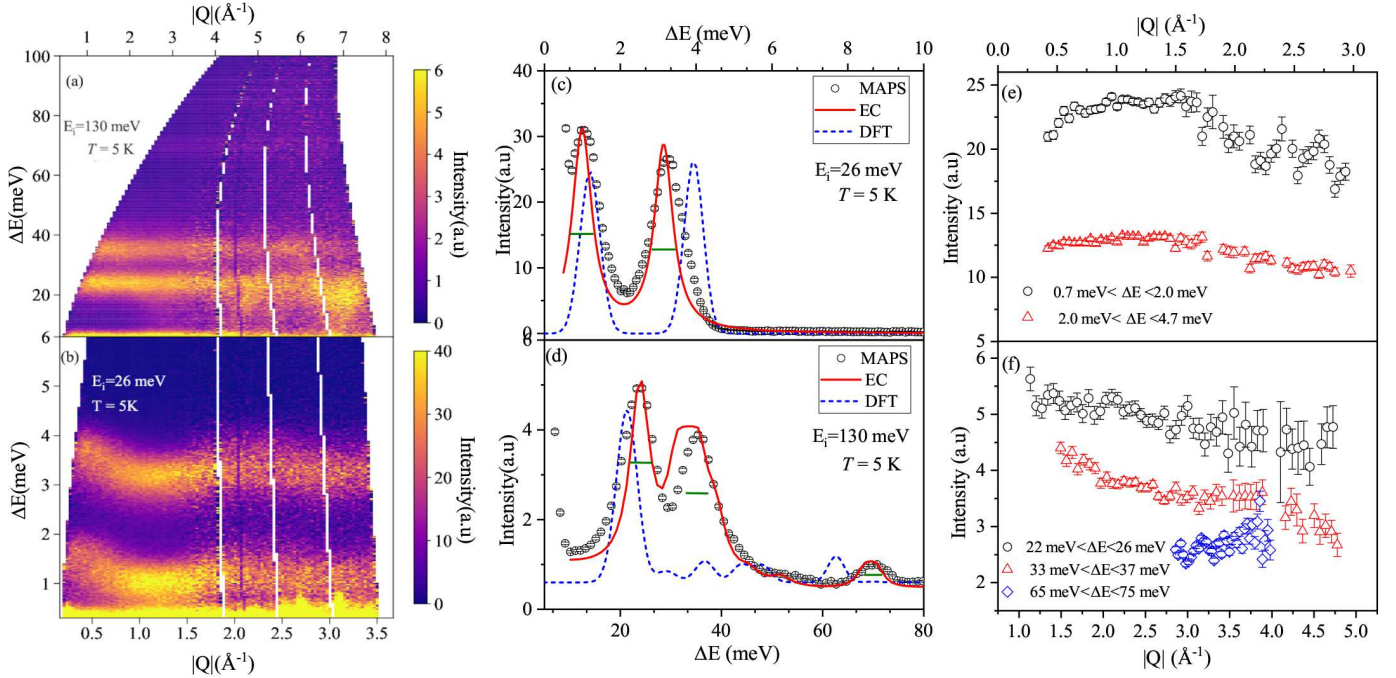}
         \caption{
				(Color online) INS spectrum of powder \stm \ as a function of momentum transfer ($\left|{\rm Q}\right|$) and energy transfer (${\rm \Delta E}$) with (a) E$_i=130$\,meV and (b) E$_i=26$\,meV. The spectrum was recorded using the MAPS spectrometer at 5\,K. The color intensity scale indicates neutron counts, where purple stands for low counts and yellow high counts in arbitrary units. The gaps in the spectra are due to gaps between adjacent detectors. Integrated INS intensity (open symbols) as a function of energy transfer for (c) E$_i = 26$\,meV integrated in the range 0.7\,\AA$^{-1}<Q<2.5$\,\AA$^{-1}$ and (d) E$_i$ = 130\,meV integrated in the range 1\,\AA$^{-1}<Q<4$\,\AA$^{-1}$. Simulated spectra using the DFT  (blue dashed line) and EC models (red solid line). Green lines refer to the calculated instrument resolution at the specific energy transfer \cite{Mantid}. $\left|{\rm Q}\right|$ dependency of the identified modes in the spectra for (e) E$_i = 26$\,meV  and (f) E$_i= 130$\,meV. }
    \label{FIG:Tof}
\end{figure*}

\section{Methods}
Powder \stm\, samples were prepared according to Ref. \cite{Karunadasa2005}, and single crystals were synthesized as described in Ref.\cite{Li2015}. A single crystal sample of mass $2.9$\,mg was used to measure the magnetization in the temperature range of $2–300$\:K with a magnetic field of $0.1$\:T applied along with main crystallographic directions. Heat capacity measurements were performed in a single-crystal sample of mass $49$\:mg in the temperature range of $2-200$\:K. These measurements were performed using a Physical Property Measurement System.

High-field magnetization was measured in a $^4$He-flow cryostat up to $58$\,T using a coaxial pickup coil system \cite{Skourski2011}. The calibration to the absolute value of magnetization was done using a continuous field magnetometer in a commercial Superconducting Quantum Interference Device.

INS experiments were performed on a polycrystalline sample of mass 3.9\,g at 5\,K using the direct geometry neutron Time-of-Flight (TOF) instrument MAPS at ISIS \cite{Ewings2019}. The MAPS spectrometer was operated in two multi-rep modes yielding two incident energies ($E_i$) 26\,meV and 130\,meV using the sloppy chopper operated at 400\,Hz.

The DFT model uses DFT derived Wannier functions, which takes the anisotropic shape features of \emph{f}-orbitals into consideration \cite{Novak2013,Scaramucci2015}.  The implementation procedure of this model for the APW+lo program, \texttt{WIEN2k}, was provided by Nov{\'a}k, $et.al$ \cite{ Novak2013, Blaha2020}. This procedure consists of two steps. In the first step, the standard self-consistent solution of the Kohn-Sham equations of the DFT was performed with \texttt{WIEN2k} \cite{Blaha2020}. The PBE-GGA exchange-correlation functional was used for this calculation. The atomic sphere radii for Tm$= 2.4$\,a.u, Sr $= 2.24$\,a.u and O $= 1.6$\,a.u were chosen. The number of basis sets amounted to $\sim 4023$ (corresponding to $RKmax = 7.0$), the number of $k$ points in the irreducible part of the Brillouin zone was 200 ($= 5 \times 4 \times 10$). In this step, the \tm (4\emph{f}) states were considered as the core states (open core) to avoid non-physical self-interactions that would dominate the crystal field Hamiltonian.  In the following step, \emph{f}-electrons of one of the inequivalent \tm were treated as valance states to allow their hybridization with the ligand orbitals (oxygen $2s$ and $2p$). The relative position of $4$\emph{f} and ligand orbitals were adjusted by introducing a correction term $\Delta$, which approximates the charge transfer energy. The non-self-consistent calculation was then performed, which yields the $4$\emph{f} Bloch states. These were transformed into Wannier functions using \texttt{Wannier90} and \texttt{Wien2wannier} \cite{Mostofi2014, Kunes2010}. This calculation yields the $4$\emph{f} local Hamiltonian, and it was then expanded in spherical tensor operators to obtain the CFPs, which are the expansion coefficients in Wybourne normalization \cite{Novak2013, Wybourne1965}. Series of calculations with different $\Delta$ values in the range $-0.68$\,eV to $-13.6$\,eV were performed and compared with experimental results such as INS levels and bulk properties. The value of $\Delta =-1.83$\,eV was found to emulate the experimental results closely.

The EC model is based on the crystallographic PC model where the charge on oxygen ligands surrounding \tm \ are considered as effective point charges lying between Tm-O bonds at effective radii. The starting values of effective charges are estimated by considering a net charge contribution to the oxygen by three \tm and two Sr$^{2+}$ coordinating with every oxygen.  Subsequently, optimal effective charge and positions are obtained by fitting INS data using the code provided with Ref. \cite{Dun2021} and \texttt{SIMPRE} \cite{Baldovi2013}. The DFT model provides approximate estimations of crystal field levels that originate from Tm1 and Tm2. These estimations were considered to perform the fitting efficiently. Effective charges $-0.468e$ and $-0.725e$ and effective radii of 1.571\,\AA\, and 1.736\,\AA\ are obtained from fitting INS data for Tm1 and Tm2 respectively. The CFP's for DFT model and EC model in Stevens normalization are presented in table \ref{tab:table1}.

Zero-field (ZF-) and longitudinal field (LF-) muon spin resonance \mpsr \ measurements were performed with a polycrystalline sample, using the instruments DOLLY and LTF at the Swiss muon source (S$\mu$S), PSI. For the experiments, 2\,g of polycrystalline sample was mixed with alcohol diluted GE varnish and attached to a silver plate for good thermalization. The LTF instrument with a dilution refrigerator was used to make two zero-field measurements at 19\,mK and 1.4\,K.  ZF- measurements at several temperatures (between  2-250\,K) were made on the DOLLY instrument. In addition, LF- measurements were made at 70\,K and 5\,K  with several magnetic fields applied in the range 0.05\,mT-0.5\,T using DOLLY.

To understand the muon-induced effects, theoretical calculations ($\mu^+$-DFT) were used to estimate the muon stopping site and the impact of muon on the local environment. The calculations were performed with the plane wave pseudopotential program \texttt{QUANTUM ESPRESSO} \cite{Giannozzi2009} using GGA exchange-correlation. Ions were modeled using ultrasoft pseudopotentials, and the muon was modeled by a norm-conserving hydrogen pseudopotential following the calculations in Ref. \cite{Foronda2015}.  
The calculations were performed in a single unit cell. The \texttt{Mufinder} was used to assign the starting positions for muons \cite{Huddart2020}. The system was allowed to relax until all the forces were below $10^{-3}$\,Ry/a.u. The calculations were performed with sets of several lowest inter-muon and muon to atom distances as constraints. Effect of muon implantation on crystal fields were then quantified using Hutchings PC model by comparing crystal fields before and after muon implantation.

\begin{table}[b]
\caption{\label{tab:table1}
CFPs $B_l^{m}$(\,meV) in Stevens  normalization for DFT model and EC model.}
\resizebox{0.43\textwidth}{!}{\begin{tabular}{|c||c|c|c|c|}
\hline \hline $B_l^{m}$	&	DFT Tm1	&	DFT Tm2	&	EC-Tm1	&	EC-Tm2 
 \tabularnewline \hline \hline
$B_2^{0}	$&$	0.086866	$&$	-0.100362	$&$	0.118755	$&$	0.075761	$	\tabularnewline	
$B_2^{2}	$&$	0.344579	$&$	-0.553521	$&$	-0.06381	$&$	-0.253117	$	\tabularnewline	
$B_2^{-2}	$&$	0.114204	$&$	-0.576164	$&$	0.33965	$&$	0.477003	$	\tabularnewline	
$B_4^{0}	$&$	-0.000518	$&$	-0.000558	$&$	-0.003498	$&$	-0.002377	$	\tabularnewline	
$B_4^{2}	$&$	-0.018459	$&$	0.011757	$&$	-0.024412	$&$	0.0313	$	\tabularnewline	
$B_4^{-2}	$&$	-0.0184	$&$	0.014482	$&$	-0.011755	$&$	0.008587	$	\tabularnewline	
$B_4^{4}	$&$	-0.004236	$&$	0.00228	$&$	-0.000164	$&$	0.005889	$	\tabularnewline	
$B_4^{-4}	$&$	0.022274	$&$	0.016151	$&$	0.031696	$&$	0.016069	$	\tabularnewline	
$B_6^{0}	$&$	0.000046	$&$	0.000042	$&$	0.000048	$&$	0.000049	$	\tabularnewline	
$B_6^{2}	$&$	0.000152	$&$	0.000075	$&$	0.000001	$&$	0.000003	$	\tabularnewline	
$B_6^{-2}	$&$	0.000136	$&$	-0.000075	$&$	-0.000002	$&$	-0.000015	$	\tabularnewline	
$B_6^{4}	$&$	-0.000044	$&$	0.000105	$&$	-0.000043	$&$	0.000026	$	\tabularnewline	
$B_6^{-4}	$&$	-0.000138	$&$	-0.000061	$&$	-0.000131	$&$	-0.000146	$	\tabularnewline	
$B_6^{6}	$&$	0.000277	$&$	-0.000118	$&$	0.00016	$&$	0.000029	$	\tabularnewline	
$B_6^{-6}	$&$	-0.000083	$&$	0.00022	$&$	-0.000111	$&$	0.00004	$	\tabularnewline	\hline

 \hline
\end{tabular}}
\end{table}

\section{Results}

\subsection{Crystal Field Analysis}

Magnetization results obtained from pulsed-field measurements at $T=1.5$\,K are compared with net magnetization obtained from DFT  and EC models in Fig. \ref{FIG:Bulk}(b).  Magnetization with fields applied along the $a$- and $b$-axes increase quickly ($<10$\,T) and slow down eventually at higher fields. The $c$-axis remains the hard axis at all fields. The theoretical saturation moment for \tm is $\SI{7}{\micro_{\rm{B}}}$. However, magnetization along $a$-direction only reaches a maximum of $\sim \SI{6.24}{\micro_{\rm{B}}}$ for maximum applied field. The measured slopes suggest that the sample is far from reaching magnetic saturation. 

The magnetization for each crystal field model was derived by taking the average of magnetization of the two inequivalent sites. The EC model predicts magnetization trends with a slight deviation. This slight mismatch could be corrected by including the Weiss-molecular fields, implying significant contributions from magnetic interactions (not done here). The net magnetization from the DFT model does not predict the magnetization trends correctly. 

A visual representation of g-tensor obtained from EC model for Tm1 and Tm2 is presented in Fig. \ref{FIG:Bulk}(c). EC model predicts easy-axis anisotropy for Tm1 and easy-plane anisotropy ($ab$-plane) for Tm2. The model predicts the $c$-axis to be the hard axis for both Tm1 and Tm2. The DFT model also predicts similar anisotropy trends however these are not shown here. Similar distinct single ion anisotropy between inequivalent sites is a common occurrence in \sln\,family \cite{Fennell2014,Malkin2015}.

Low-temperature heat capacity is presented in Fig. \ref{FIG:Bulk}(d). A broad bump observed at low temperatures is a clear sign of a Schottky anomaly, implying the existence of low energy excitations. Above 20\,K, phonon contributions to heat capacity get significantly stronger, making it difficult to distinguish any other magnetic contributions.  Both crystal field models predict the Schottky anomaly fairly accurately indicating that the models predict the low energy excitations accurately.

Measured and calculated magnetic susceptibility, are shown in figures \ref{FIG:Bulk}(e,f). These experimental results are a good indicator of the high magnetic anisotropy in this compound. The susceptibility along the $c$-axis is an order of magnitude smaller than for other directions, again reiterating that it is the hardest axis for magnetization. The susceptibility follows a paramagnetic behavior for temperatures below 50\,K, followed by a broad peak centered at 220\,K.  On the other hand, the susceptibility for fields applied along the $a$- and $b$- axes follow a Curie-Weiss behavior all the way down to 50\,K, where the curves start leveling off. None of the curves show any sign of a transition to a long range magnetic order. The EC model predicts susceptibility trends significantly well at high temperatures. At low temperatures however, it deviates from the experimental results. This could be due to unaccounted contributions from magnetic interactions.

\begin{table*}[h]
    \begin{subtable}[h]{0.5\textwidth}
        \centering
      \resizebox{1\textwidth}{!}{\begin{tabular}{|c||c|cccccccc|}
\hline \hline Model & E (\,meV) &$|0\rangle$ & $|\pm 1\rangle$ & $|\pm 2\rangle$ & $|\pm 3\rangle$ & $|\pm 4\rangle$ & $|\pm 5\rangle$ & $|\pm 6\rangle$ &\tabularnewline
 \hline \hline
     & 0.000 &  0.695   &   &    0.0586    &   &   0.079      &  &    0.015  &      \tabularnewline
EC     & 3.143  &  & 0.4333       &  &   0.0295     & &   0.0371      &  &    \tabularnewline 
   & 23.162  & &  0.2411     &  &   0.1586    &  &   0.1003     &  &     \tabularnewline 
    & 30.492  &   0.0629     &  &  0.4041      &  &   0.0133      &  &   0.0512    &         \tabularnewline\hline \hline
   &  0.000  & 0.5688 &        & 0.1108 &        & 0.0927  &         & 0.0041 &      \tabularnewline
DFT    & 3.935  &        &  0.4467&        & 0.0056 &         & 0.0408  &        &      \tabularnewline 
    & 21.608  &        & 0.0589 &        & 0.1849 &         & 0.2488  &       &       \tabularnewline
     & 35.589  &        & 0.0258 &        & 0.1098 &         & 0.3591  &      &           \tabularnewline \hline \hline
PC & 0.000  & 0.6083&        & 0.1577&        & 0.0301&          &         &   \tabularnewline
    & 1.228 &         & 0.455&       & 0.0281&        & 0.011  &          & \tabularnewline \hline \hline
$\mu^+$-PC  & 0.0000 & 0.4896 &  & 0.2133 &  & 0.0347 & &  &  \tabularnewline
             & 0.234 &  & 0.3951 &  & 0.0902 &  & 0.0087 &  & \tabularnewline \hline \hline
\end{tabular}}
       \caption{Tm1}
       \label{tab:table4}
    \end{subtable}
 %   \hfill
    \begin{subtable}[h]{0.5\textwidth}
        \centering
\resizebox{1\textwidth}{!}{\begin{tabular}{|c||c|cccccccc|}
\hline \hline Model & E (\,meV) &$|0\rangle$ & $|\pm 1\rangle$ & $|\pm 2\rangle$ & $|\pm 3\rangle$ & $|\pm 4\rangle$ & $|\pm 5\rangle$ & $|\pm 6\rangle$ &\tabularnewline
 \hline \hline
       & 0.000  &   0.5597 &   &  0.1532      & &    0.0499     &  & 0.017     &       \tabularnewline 
EC    & 0.994   &  &  0.4029       &  &   0.0667     &  &  0.0304      & &       \tabularnewline 
    & 32.997  &  0.0048     &  &   0.0355     &  &  0.1898       &   &  0.2724    &       \tabularnewline 
    & 36.585  &  &  0.0256        & &   0.1154    &   &   0.359      &  &    \tabularnewline \hline \hline 
    & 0.000  &        &  0.4000 &        & 0.0006 &         & 0.0884  &       &       \tabularnewline 
DFT    & 1.202   & 0.6537 &         & 0.13   &        & 0.0339  &         & 0.0027&       \tabularnewline
    & 20.387  &        & 0.0589  &        & 0.1849 &         & 0.2842  &       &       \tabularnewline
    & 21.379  & 0.0076 &         & 0.1747 &        & 0.1508  &         & 0.1652 &    \tabularnewline \hline \hline
 PC  & 0.000 & 0.4423 &  & 0.2163 & & 0.0508 &  &  & \tabularnewline
      & 0.388 &  & 0.369 &  & 0.107 & & 0.0183 &  & \tabularnewline \hline 
 $\mu^+$-PC  & 0.0000 & 0.212 &  & 0.1630 & &0.0983&  &  &\tabularnewline
              & 0.24 &  & 0.1965 &  & 0.1311 & & 0.0574&  & \tabularnewline \hline \hline
\end{tabular}}
        \caption{Tm2}
        \label{tab:table5}
     \end{subtable}
     \caption{Eigenvectors and Eigenvalue crystal field models.}
     \label{tab:temps}
\end{table*}

Powder INS spectra measured at 5\,K with incident energies $26$ and $130$\,meV are presented as a function of momentum transfer ($\left|{\rm Q}\right|$) and energy transfer ($\Delta{\rm E}$) in figures \ref{FIG:Tof}(a, b). The specific incident energies were chosen to cover the entire energy range in which the ground state multiplet expands (${\rm E}<110$\,meV).  Integrated neutron scattering intensity as a function of energy transfer is presented in figures \ref{FIG:Tof}(c, d). In the low energy excitation spectra, two dispersing excitations can be identified centered at 1.230(5)\,meV and 3.412(3)\,meV with minima at $\sim 1.2$\,\AA$^{-1}$. In the higher energy range, three excitations centered at 23.72(4)\,meV, 35.82(6)\,meV and 71.00(2)\,meV are identified. The expected instrument resolution was calculated using \texttt{Mantid} \cite{Mantid} for each of the centers of the identified peaks. These are shown as green horizontal lines in figures \ref{FIG:Tof}(c,d). It can be noted that the all peaks are slightly broader than the resolution. This is expected for the lowest two modes, as they are dispersing. No dispersion can be seen however for the higher modes, and the slight broadening could be due to the overlapping of multiple excitations. Integrated intensity as a function of momentum transfer for each of these identified modes is plotted in figures \ref{FIG:Tof}(e,f). The decreasing intensity for increasing values of $\left|{\rm Q}\right|$ \ is an indicator of the magnetic nature of the excitations. It is important to note that the low energy modes (Fig. \ref{FIG:Tof}(e)) are influenced by the dynamic structure factor due to inter-ionic magnetic interactions, showing a maximum centered at 1.2\AA$^{-1}$. The $\left|{\rm Q}\right|$ of mode centered at 71\,meV, does not show the typical magnetic form factor. This could be due to strong phonon contributions at the measured restricted $\left|{\rm Q}\right|$ range and energy transferred. 

The fitted and calculated neutron powder scattering cross-section for EC and  DFT models are presented in figures \ref{FIG:Tof}(c,d). The cross-sections were convoluted with the MAPS instrument resolution function using \texttt{Mantid} \cite{Mantid}. The EC model fits INS data well. A small disagreement is found around $30.492$\,meV, where the model calculates additional crystal field modes. Contrarily, the DFT model is successful in explaining only the low-energy excitations. 

The first four eigenvalues and eigenvectors are reported in tables \ref{tab:table4} and \ref{tab:table5} for the DFT and EC models.  $J$-mixing was observed within $|^{3}H_{6},m_J\rangle$. The EC model predicts major contributions from $J = 0$ to the ground states of both the sites and $J=\pm 1$ dominates the first excited states. In contrast, in the DFT model, the major contribution to the Tm1 ground state comes from $J = 0$, while the ground state of Tm2 $J = \pm 1$ provides a major contribution. It is evident that among the low energy excitations, the first excited state originates from Tm2 while the second excited state comes from Tm1.

\subsection{\mpsr \ - Muon Spin Rotation and Relaxation}

\begin{figure*}[htb!]
\centering
        \includegraphics[width=1\textwidth]{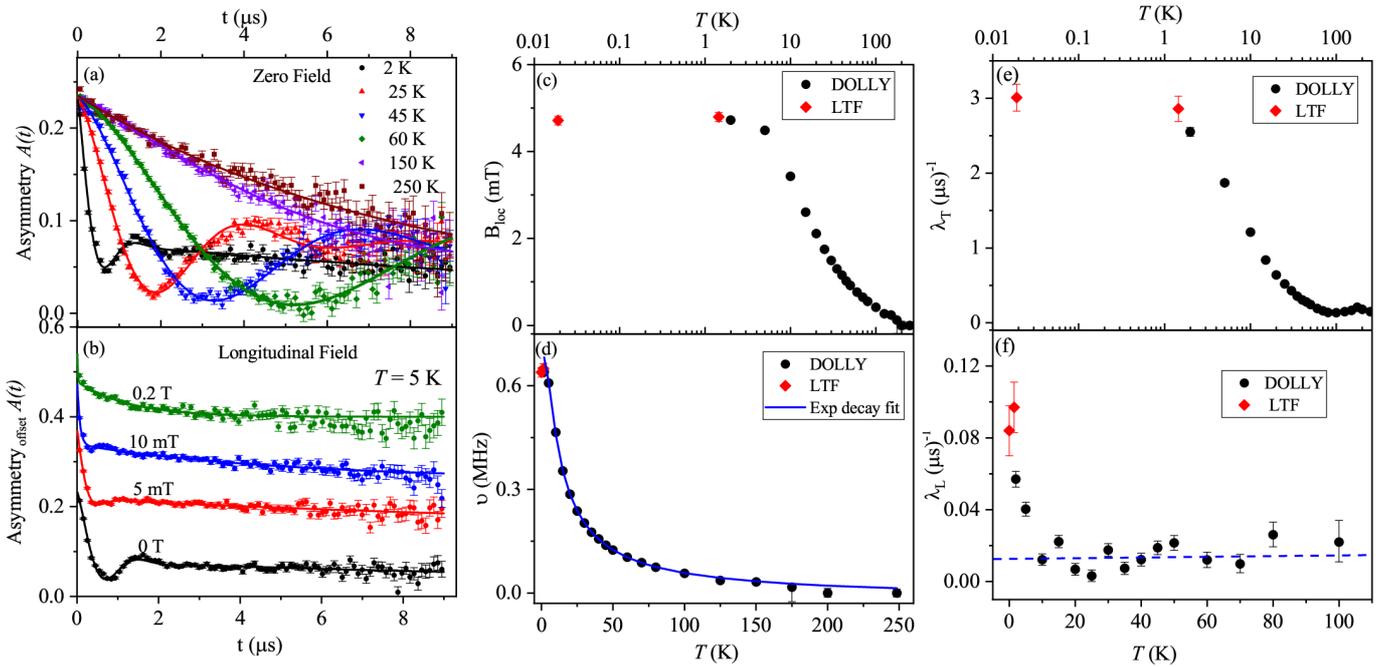}
\caption{(Color online)  
(a) ZF-\mpsr \ spectra at selected temperatures. The solid lines represent the model described in Eq. (\ref{eq:muSR1}). (b) LF-\mpsr \ spectra at selected applied magnetic fields at 5\,K. 
(c) Local field as a function of temperature extracted from the fitting to the ZF-\mpsr \ data. (d) Muon precession frequency and exponential decay fit. Extracted transverse (e) and longitudinal (f) relaxation rates. The dashed line at $0.01(\mu s)^{-1}$ in (f) is a guide to the eye.
 }
\label{Fig:muonALL}
\end{figure*}

ZF-\mpsr \  asymmetry for selected temperatures are presented in Fig. \ref{Fig:muonALL}(a). At low temperatures ($T<100$\,K), along with the relaxation part, there are oscillations with a single frequency that start developing and damps as the time evolves. At higher temperatures, these oscillations move beyond the instrument's time window, and only the relaxing part is available. The oscillations in the data are due to the muon precession as the muon experiences a static magnetic field in its immediate vicinity. Such oscillations are usually linked to long-range magnetic order, and a single frequency observation is the evidence for a single muon stopping site \cite{Yaouanc2011}. Despite this, there is no other evidence of long-range order in this compound, as discussed before.  

The  \mpsr \ spectra were modeled using the polarization Eq. (\ref{eq:muSR1})  for the entire temperature range (19\,mK-250\,K) \cite{Mulders1998}.

\begin{equation}\label{eq:muSR1}
 {a_0P_Z^{exp}(t)}= a_0[\frac{1}{3}e^{-\lambda_L t}+\frac{2}{3}e^{(-\lambda_T t)} cos(2\pi\nu t)]
\end{equation}

Where $a_0$ is the initial asymmetry and $P_Z^{exp}(t)$ the polarization and $\nu$ is the muon precession frequency. $\lambda_L$ (L= longitudinal) and $\lambda_T$ (T = transverse) are the relaxation rates parallel and perpendicular to the local field. In polycrystalline samples, the ratio of amplitudes between parallel and perpendicular components is $1:2$ (fixed in Eq. (\ref{eq:muSR1})) \cite{Cooke1995}. The mean value of local field experienced by a muon at its stopping site $B_{loc}$  can be obtained from the Larmor precession frequency using $2\pi\nu = \gamma_\mu B_{loc}$, where $\gamma_\mu$ is the gyromagnetic ratio of the muon ($0.1355$\,MHz/mT) \cite{Yaouanc2011}. 

In polycrystalline samples, one can estimate $\lambda_L$ by observing a $1/3$ tail of the asymmetry spectra and $\lambda_T$ a $2/3$ tail. At low temperatures, the relaxation of both $1/3$ and $2/3$ tails is clearly visible. However, above 45\,K, the 1/3 tail slowly shifts beyond 9\,$\mu$s, the upper limit of the instrument, and vanishes completely above 100\,K. Parts of the $2/3$ tail are not visible beyond 150\,K. Hence our understanding of temperature dependence of ZF- relaxation rates and muon precession frequency is limited by the instrument time window.  

The muon precession frequency $\nu$ and the $B_{loc}$ are presented in figure \ref{Fig:muonALL}(c,d). It is important to note that in a compound exhibiting no spontaneous magnetic order, $B_{loc}$ is only different to zero in an applied magnetic field. Here, however, $B_{loc}$ increases exponentially while cooling down, leveling off for temperatures below $2$\,K at $4.7$\,mT. There is no clear critical temperature, unlike any standard second-order phase transition to an ordered state. The muon precession frequency follows a similar trend, its temperature dependency has been fitted to an exponential decay function of the form $exp (-\delta /T)$\, \cite{Pregelj2012}, where $\delta=11.3(3)$\,K ($0.97$\,meV), seen as a blue line in Fig.\ref{Fig:muonALL}(d). This activation temperature, $\delta$, has coincidentally the same value of the gap to the first excited state.

The transverse and longitudinal muon spin relaxation rates extracted from the fits are presented in figures \ref{Fig:muonALL}(e, f). $\lambda_T$ is at least an order of magnitude faster than $\lambda_L$ and has a similar temperature dependence as that of $\nu$. However, $\lambda_L$ remains almost constant, i.e., $0.01(\mu s)^{-1}$ above 10\,K, and sharply increases below. 
The origin of the transverse component of the relaxation rate could either be due to the distribution of the local field or the dephasing of muon precession from fluctuation effects \cite{Cooke1995}. On the other hand, the longitudinal component is purely due to fluctuations (dynamic effects). If the transverse component of the relaxation was due to static or quasi-static field distribution, then the field distribution should have narrowed as the local field (ordering parameter) reaches saturation. Nevertheless, we see that the relaxation rate increases as the local field increases saturating at 2\,K.  This behavior is a sign of an extended critical regime,  similar to the one reported for TmNi$_2$B$_2$C \cite{Cooke1995}. 

Also, a consistent observation can be made in the case of the longitudinal field measurements. Fig. \ref{Fig:muonALL}(b) shows the LF-\mpsr \ time evolution spectra captured at $T=5$\,K  with various applied magnetic fields in longitudinal geometry. This technique is used to determine whether the damping of muon polarization is caused due to the distribution of static fields or relaxation due to fluctuations \cite{Yaouanc2011}. The asymmetry spectra measured at several fields show that the oscillations are quenched at an applied field of $5$\,mT, which is close to $B_{loc}$ value at 5\,K. Nonetheless, there is no sign of relaxation quenching even at $0.2$\,T. 

\begin{figure}[htb!]
\centering
        \includegraphics[width=0.5\textwidth]{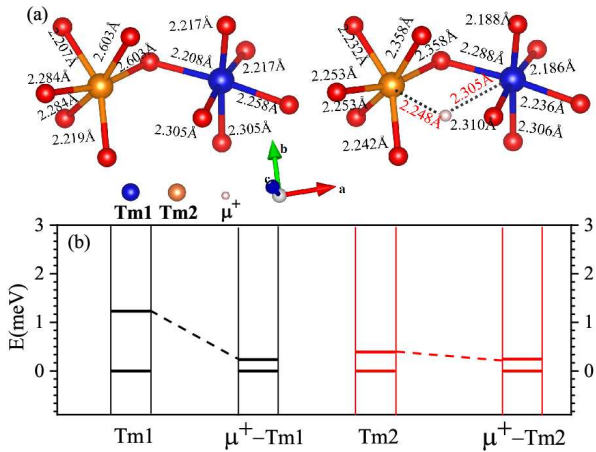}
         \caption{(Color online) (a) Bond lengths comparisons before (extracted from Ref.\cite{Li2015}) and after muon implantation. (b) Impact of the muon-induced distortion in the crystal field scheme calculated using the PC model. }
    \label{Fig:muon3}
\end{figure}

The muon stopping site was estimated using DFT based structural relaxation calculation, as explained in the methods section. A plane-wave pseudopotential program was used as it is relatively faster than all-electron DFT programs for structural relaxation. The presence of a single muon stopping site was considered as a single precession frequency was observed in the ZF measurements. The calculation yields a low symmetry interstitial site, as shown in Fig. \ref{Fig:muon3}(a). The effects of muon implantation on the crystal field scheme were modeled using the PC model and are shown in Fig. \ref{Fig:muon3}(b). Although the PC model is not accurate, it gives a qualitative distinction between the unperturbed and muon-perturbed crystal fields. The muon-induced distortion reduces the gap to the first excited state by $\sim$ 80$\%$\ for Tm1 and by  $\sim$ 38$\%$\ for Tm2. 

\section{Conclusions}

The crystal field properties of the non-Kramers compound \stm \ were investigated using a combination of susceptibility, magnetization in pulsed fields, heat capacity, INS, and polarized muon spectroscopy. Two crystal field models, the DFT model, and the EC model are proposed to understand the single-ion properties. The DFT model foresees the ground state dominated by $J = 0$ for Tm1 and $J=\pm 1$ for Tm2 while the EC model determines the ground state of both the inequivalent sites to show predominantly $J = 0$ characteristics. Both the models predict easy-axis anisotropy for Tm1 and easy-plane anisotropy for Tm2. The EC model predicts the magnetic properties to great extent despite some small discrepancies at low temperatures, which can be attributed to inter-ionic magnetic interactions. These will be discussed in a separate report.

On the other hand, the DFT model is far from being perfect. Accuracy of the DFT model CFPs are strongly determined by the accurate determination of $\Delta$. Although \texttt{Wannier90} program calculates MWLFs, it does not guarantee that they will be centered on the crystallographic site of the \tm \ ions. To avoid this, calculations with other Wannier function formalism could be implemented as suggested in Ref. \cite{Novak2015}. Additionally, even though open core calculations should restrain the $4f$ electrons in the core, some of their density can still leak out. Despite these shortcomings, the DFT model has provided a preliminary understanding of the crystal field schemes that have helped in making efficient fitting of INS spectra to arrive at the EC model. 

ZF-\mpsr \ results show oscillations in the asymmetry spectra, a standard signature of long-range order. However, the absence of standard critical behavior and absence of long or short-range order reported in Ref. \cite{Li2015} indicates that the observed ordering is a muon induced phenomenon. To quantify the impact of muon-induced distortion on the crystal field, the muon stopping site was determined using DFT techniques, and then the crystallographic PC model was used to estimate the crystal field. The PC model qualitatively determines how muon implantation can renormalize the gap size to the low energy crystal field levels by up to $80\%$ on Tm1 and $38\%$ on Tm2. The renormalized gaps would take a value of $\approx 0.6$\,meV ($\approx 7$\,K) for both ions. Having eventually an effective pseudo-doublet ground state. Similar observations have been made in the case of other non-magnetic non-Kramers ion-based systems \cite{Kayzel1994,Foronda2015,Feyerherm1995} where it has been concluded that the observation of local fields was due to a muon-induced perturbation in the crystal field scheme. In these cases, however, hyperfine interactions play an important role. 

Further insights are revealed by analyzing the temperature dependence of the muon precession frequency. The muon precession frequency follows an exponential decay with a thermal activation gap of $11.3(3)$\,K, which is the same as the gap to the first excited state measured with INS. This together with the temperature dependence of the longitudinal field measurements allow us to conclude that the observed relaxation is dynamic in origin and that the local field is rapidly fluctuating, making the muon experience a quasi-static local field.
%%%%%%%%%%%%%%%%%%%%%%%%%%%%%%%%%%%%%%%%%%%%%%%%%%%%%%%%%%%%%%%%%%%%%%%%%%%%%%%%%%%%%%%%%%%%%%%%%%%%%%%%%%%%%%%%%%%%%%%%%%%%%%%%%%%%%%%%%%%%%%%%%%%%%%%%%%%%%%%%%%%%%%%%%%%%%%%%%%%%%%%%%%%%%%%%%%%%%%%%%%%%%%%%%%%%%%%%%%%%%%%%%%%%%
\section{Acknowledgments}

We acknowledge valuable discussions with G. Nielsen, P. Nov{\'a}k, and M. D. Le. Experiments at the ISIS Pulsed Neutron and Muon Source were supported by a beamtime allocation from the Science and Technology Facilities Council. We acknowledge the support of HLD at HZDR, a member of European Magnetic Field Laboratory (EMFL). This work is based on experiments performed at the Swiss Muon Source S$\mu$S Paul Scherrer Institute, Villigen, Switzerland. We would like to acknowledge the support of the CoreLab Quantum Materials at Helmholtz-Zentrum Berlin. We want to thank all the facilities for their kind support.

%%%%%%%%%%%%%%%%%%%%%%%%%%%%%%%%%%%%%%%%%%%%%%%%%%%%%%%%%%%%%%%%%%%%%%%%%%%%%%%%%%%%%%%%%%%%%%%%%%%%%%%%%%%%%%%%%%%%%%%%%%%%%%%%%%%%%%%%%%%%%%%%%%%%%%%%%%%%%%%%%%%%%%%%%%%%%%%%%%%%%%%%%%%%%%%%%%%%%%%%%%%%%%%%%%%%%%%%%%%%%%%%%%%%%
%\online{DFTWF_1},\cite{DFTWF_3},\cite{DFTWF_4},
\bibliography{SrTm2O4_JMMM}

\end{document}